\documentclass[a4paper,11pt]{article}
\usepackage{jheppub} 
\usepackage[mathscr]{euscript}
\usepackage{multirow,stmaryrd}
\numberwithin{equation}{section}

\preprint{DESY-23-073}

\title{Probing M-theory with tetrahedron instantons}

\author[1]{Elli Pomoni,}
\author[2]{Wenbin Yan,}
\author[1,3]{and Xinyu Zhang}

\affiliation[1]{Deutsches Elektronen-Synchrotron DESY, Notkestr. 85, 22607 Hamburg, Germany}
\affiliation[2]{Yau Mathematical Sciences Center, Tsinghua University, Shuangqing Road, Beijing, 100084, China}
\affiliation[3]{Zhejiang Institute of Modern Physics, School of Physics, Zhejiang University, 866 Yuhangtang Rd, Hangzhou, Zhejiang, 310058, China}

\emailAdd{elli.pomoni@desy.de}
\emailAdd{wbyan@tsinghua.edu.cn}
\emailAdd{xinyu.zhang@zju.edu.cn}

\abstract{
The duality between type IIA superstring theory and M-theory enables
us to lift bound states of D$0$-branes and $n$ parallel D$6$-branes
to M-theory compactified on an $n$-centered multi-Taub-NUT space
$\mathbb{TN}_{n}$. Accordingly, the rank $n$ K-theoretic Donaldson-Thomas
invariants of $\mathbb{C}^{3}$ are connected with the index of M-theory
on $\mathbb{C}^{3}\times\mathbb{TN}_{n}$. In this paper, we extend
this connection by considering intersecting D$6$-branes. In the presence
of a suitable Neveu-Schwarz $B$-field, the system preserves two supercharges.
This system is T-dual to the configuration of tetrahedron instantons
which we introduced in \cite{Pomoni:2021hkn}. We conjecture a closed-form
expression for the K-theoretic tetrahedron instanton partition function,
which is the generating function of the D$0$-D$6$ partition functions.
We find that the tetrahedron instanton partition function coincides
with the partition function of the magnificent four model for special
values of the parameters, leading us to conjecture that our system
of intersecting D$6$-branes can be obtained from the annihilation
of D$8$-branes and anti-D$8$-branes. Remarkably, the K-theoretic
tetrahedron instanton partition function allows an interpretation
in terms of the index of M-theory on a noncompact Calabi-Yau fivefold
which is related to a superposition of Kaluza-Klein monopoles. The
dimensional reduction of the system allows us to express the cohomological
tetrahedron instanton partition function in terms of the MacMahon
function, generalizing the correspondence between Gromov-Witten invariants
and Donaldson-Thomas invariants for Calabi-Yau threefolds.
}


\arxivnumber{2306.06005}

\begin{document}

\maketitle


\section{Introduction}

One of the most striking consequences of string duality is the discovery
of an eleven-dimensional interacting quantum theory known as M-theory
\cite{Townsend:1995kk,Witten:1995ex}, whose low-energy limit is the
eleven-dimensional supergravity \cite{Cremmer:1978km}. Despite the
lack of an intrinsic formulation of M-theory in terms of its fundamental
degrees of freedom, there has been diverse and convincing evidence
of its existence. Especially, ten-dimensional type IIA superstring
theory at finite string coupling is conjectured to be equivalent to
M-theory compactified on a circle. All objects in type IIA superstring
theory, including fundamental strings, D-branes, and NS$5$-branes,
can be naturally lifted in M-theory. 

Over the past few decades, the conjectured duality between type IIA
superstring theory and M-theory has led to many interesting results
by comparing the computations of BPS quantities from two viewpoints.
One prominent setup is the system of D$0$-branes probing $n$ parallel
D$6$-branes in type IIA superstring theory on $\mathbb{S}_{t}^{1}\times\mathbb{C}^{3}\times\mathbb{R}^{3}$
\cite{maulik2006gromov1,maulik2006gromov2,Cirafici:2008sn}. This
configuration is supersymmetric if a suitable Neveu-Schwarz $B$-field
is turned on \cite{Witten:2000mf}. The generating function of D$0$-D$6$
partition functions with varying number of D$0$-branes can be interpreted
as the instanton partition function of the $\left(6+1\right)$-dimensional
noncommutative $\mathrm{U}(n)$ super Yang-Mills theory on $\mathbb{S}_{t}^{1}\times\mathbb{C}^{3}$
\cite{Baulieu:1997nj,nekrasov2009instanton}, and can be computed
exactly \cite{Moore:1998et}. Mathematically, this instanton partition
function computes rank $n$ K-theoretic Donaldson-Thomas invariants
of $\mathbb{C}^{3}$ \cite{Donaldson:1996kp,Nekrasov:2014nea,Okounkov:2015spn}.
Upon lifting to M-theory, the D$0$-branes become Kaluza-Klein (KK)
modes of the graviton, and the D$6$-branes become KK monopoles, which
can be described geometrically by an $n$-centered multi-Taub-NUT
space $\mathbb{TN}_{n}$ \cite{Townsend:1995kk,Hull:1997kt,Sen:1997js,Burgess:2003mm,Lee:2008ha}.
This means that the instanton partition function is related to the
partition function of M-theory on $\mathbb{S}_{t}^{1}\times\mathbb{C}^{3}\times\mathbb{TN}_{n}$,
or equivalently the index of M-theory on the noncompact Calabi-Yau
fivefold $\mathbb{C}^{3}\times\mathbb{TN}_{n}$. This sharp prediction
has been confirmed for $n=1$ in \cite{nekrasov2009instanton} and
later for $n\geq2$ in \cite{Benini:2018hjy}. \footnote{There are further generalizations by replacing $\mathbb{C}^{3}$ with
other noncompact Calabi-Yau threefolds or by introducing D$2$/D$4$-branes
\cite{Nekrasov:2014nea,Zotto:2021xah,Cirafici:2021yda,Nekrasov:2021ked}.
Such generalizations are beyond the scope of this paper, and we will
concentrate on the simplest example which is already interesting enough.}

We notice that on both sides of this beautiful relation, there are
four preserved supercharges, but two of them are not used in performing
the exact computations. This observation opens an opportunity to extend
the setup by breaking half of the supercharges while still keeping
the ability to analyze exactly. We introduce additional D$6$-branes
that intersect with the original D$6$-branes in type IIA superstring
theory, producing defects of real codimension two in the worldvolume
theory of the original stack of D$6$-branes. The whole system involves
D$0$-branes and four stacks of intersecting D$6$-branes on $\mathbb{S}_{t}^{1}\times\mathbb{C}^{4}\times\mathbb{R}$.
This is in fact T-dual to the brane configuration that realizes tetrahedron
instantons, which we introduced in \cite{Pomoni:2021hkn}. In the
presence of a suitable constant Neveu-Schwarz $B$-field, two supercharges
are preserved. In \cite{Pomoni:2021hkn}, we have defined and calculated
the cohomological/K-theoretic/elliptic tetrahedron instanton partition
function. The result is expressed as a statistical sum over a collection
of random plane partitions. In this paper, we conjecture that the
K-theoretic tetrahedron instanton partition function $\mathcal{Z}_{\vec{n}}$
admits an elegant closed-form expression in terms of the plethystic
exponential (\ref{eq:ZnPE}, \ref{eq:Fgeneral}, \ref{eq:Deltan}),
\begin{equation}
\mathcal{Z}_{\vec{n}}=\mathtt{PE}\left[\frac{\prod_{1\leq a<b\leq3}\left\llbracket q_{a}q_{b}\right\rrbracket }{\prod_{a\in\underline{4}}\left\llbracket q_{a}\right\rrbracket }\frac{\left\llbracket \varDelta_{\vec{n}}\right\rrbracket }{\left\llbracket \varDelta_{\vec{n}}^{\frac{1}{2}}p\right\rrbracket \left\llbracket \varDelta_{\vec{n}}^{\frac{1}{2}}p^{-1}\right\rrbracket }\right],\quad\varDelta_{\vec{n}}=\prod_{A\in\underline{4}^{\vee}}q_{\check{A}}^{n_{A}},\label{eq:ZnIntro}
\end{equation}
where all the notations will be explained in sections \ref{sec:BraneIIA}
and \ref{sec:PartitionFunction} of the paper. Interestingly, it only
depends on the $\Omega$-deformation parameters $q_{a}$ and the instanton
counting parameter $p$, but is independent of the Coulomb parameters
associated with the positions of D$6$-branes. 

It is fascinating to connect our system with the magnificent four
model, which can be constructed using D$0$-branes probing pairs of
D8-branes and anti-D$8$-branes \cite{Nekrasov:2017cih,Nekrasov:2018xsb}.
The partition function $\mathcal{Z}^{\mathrm{MF}}$ of the magnificent
four model is given by a sum over random solid partitions. From the
mathematical viewpoint, $\mathcal{Z}^{\mathrm{MF}}$ computes the
K-theoretic version of the equivariant integral over the Hilbert scheme
of points on $\mathbb{C}^{4}$ and higher rank analogues thereof,
or equivalently Donaldson-Thomas invariants of the noncompact Calabi-Yau
fourfold $\mathbb{C}^{4}$ \cite{cao2013donaldson,cao2014donaldson,Cao:2017swr,Cao:2019tvv}.
A simple closed-form expression for $\mathcal{Z}^{\mathrm{MF}}$ was
conjectured in \cite{Nekrasov:2017cih,Nekrasov:2018xsb}. We find
that (\ref{eq:ZnIntro}) coincides with certain specialization of
$\mathcal{Z}^{\mathrm{MF}}$. This matching strongly hints that the
annihilation of D$8$-branes and anti-D$8$-branes can leave behind
a system of intersecting D$6$-branes. 

Although our system seems to be a special case of the magnificent
four model from the perspective of partition functions, it is worthy
studying independently. One important reason is that we have very
little control over the lift of D$8$-branes and anti-D$8$-branes
to M-theory, and it is utterly bewildering to provide a non-perturbative
interpretation of $\mathcal{Z}^{\mathrm{MF}}$. It was suspected in
\cite{Nekrasov:2017cih} that the magnificent four model is related
to the compactification of a mysterious thirteen-dimensional theory
$\mathscr{M}_{13}$ on a Taub-NUT space:
\begin{equation}
\begin{array}{cccccc}
\mathscr{M}_{13} & \mathbb{S}_{t}^{1} & \times & \mathbb{C}^{4} & \times & \mathbb{TN}\\
\uparrow & \parallel &  & \parallel\\
\mathrm{(anti-)D}8 & \mathbb{S}_{t}^{1} & \times & \mathbb{C}^{4}
\end{array}.
\end{equation}
However, $\mathscr{M}_{13}$ is not needed in our case since the M-theory
lift of D$6$-branes is better understood. Indeed, we find a decomposition
of $\mathcal{Z}_{\vec{n}}$, making it possible to interpret $\mathcal{Z}_{\vec{n}}$
in terms of the index of M-theory on $\mathscr{X}_{\vec{n}}$, where
$\mathscr{X}_{\vec{n}}$ is obtained by coalescing four copies of
$\mathbb{C}^{3}\times\mathbb{TN}_{n}$ into a single noncompact Calabi-Yau
fivefold. Clearly, $\mathscr{X}_{\vec{n}}$ is not a product of two
lower-dimensional manifolds, and the compactification of M-theory
on $\mathscr{X}_{\vec{n}}$ preserves two instead of four supercharges. 

We also study the dimensional reduction of our system by shrinking
the radius of $\mathbb{S}_{t}^{1}$ to zero. Accordingly, the K-theoretic
tetrahedron instanton partition function is reduced to its cohomological
counterpart. We find that the result can be expressed in terms of
the MacMahon function. This allows us to identify the tetrahedron
instanton partition function with certain gluing of A-model topological
string partition functions, generalizing the celebrated Gromov-Witten/Donaldson-Thomas
correspondence for Calabi-Yau $3$-folds \cite{maulik2006gromov1,maulik2006gromov2}. 

The rest of the paper is organized as follows. In section \ref{sec:BraneIIA},
we describe our D$0$-D$6$ brane system in type IIA superstring theory.
In section \ref{sec:PartitionFunction}, we give the definition and
the exact expression of the K-theoretic tetrahedron instanon partition
function, both in an instanton-expansion form and in a closed form.
In section \ref{sec:MF}, we discuss the connection of our system
with the magnificent four model. In section \ref{sec:M-theory}, we
explain the definition and the result of the index of M-theory. In
section \ref{sec:Relation}, we explore the relation between the K-theoretic
tetrahedron instanon partition function and the index of M-theory.
In section \ref{sec:TS}, we consider the dimensional reduction of
the system and discuss the relation with A-model topological strings,
generalizing the correspondence between Donaldson-Thomas invariants
and Gromov-Witten invariants. We end in section \ref{sec:OpenQuestions}
with a summary and a discussion of interesting open questions.

Shortly after submitting this paper to arXiv, we received a paper
by Fasola and Monavari \cite{Fasola:2023ypx} where a rigorous proof
of our conjecture (\ref{eq:ZnIntro}) was provided. They also proposed
a mathematical definition of the moduli space of tetrahedron instantons
in the language of Quot scheme of a singular variety. 

\section{D$0$-D$6$ brane system \label{sec:BraneIIA}}

\begin{table}
\begin{centering}
\begin{tabular}{|c|c|c|c|c|c|c|c|c|c|c|}
\hline 
 & $\mathbb{S}_{t}^{1}$ & \multicolumn{2}{c|}{$\mathbb{C}_{1}$} & \multicolumn{2}{c|}{$\mathbb{C}_{2}$} & \multicolumn{2}{c|}{$\mathbb{C}_{3}$} & \multicolumn{2}{c|}{$\mathbb{C}_{4}$} & $\mathbb{R}$\tabularnewline
\hline 
$x^{M}$ & $0$ & $1$ & $2$ & $3$ & $4$ & $5$ & $6$ & $7$ & $8$ & $9$\tabularnewline
\hline 
$k$ D$0$ & $-$ & $\bullet$ & $\bullet$ & $\bullet$ & $\bullet$ & $\bullet$ & $\bullet$ & $\bullet$ & $\bullet$ & $\bullet$\tabularnewline
\hline 
$n_{123}$ D$6_{123}$ & $-$ & $-$ & $-$ & $-$ & $-$ & $-$ & $-$ & $\bullet$ & $\bullet$ & $\bullet$\tabularnewline
\hline 
$n_{124}$ D$6_{124}$ & $-$ & $-$ & $-$ & $-$ & $-$ & $\bullet$ & $\bullet$ & $-$ & $-$ & $\bullet$\tabularnewline
\hline 
$n_{134}$ D$6_{134}$ & $-$ & $-$ & $-$ & $\bullet$ & $\bullet$ & $-$ & $-$ & $-$ & $-$ & $\bullet$\tabularnewline
\hline 
$n_{234}$ D$6_{234}$ & $-$ & $\bullet$ & $\bullet$ & $-$ & $-$ & $-$ & $-$ & $-$ & $-$ & $\bullet$\tabularnewline
\hline 
\end{tabular}
\par\end{centering}
\caption{Bound states of D$0$-branes and four stacks of intersecting D$6$-branes
in type IIA superstring theory. Here $-$ and $\bullet$ indicate
the worldvolume and the transverse directions of the D-branes, respectively.
\label{D0-D6}}
\end{table}

We start by describing the D$0$-D$6$ brane system in type IIA superstring
theory. This is T-dual to the string theory configuration of tetrahedron
instantons \cite{Pomoni:2021hkn}. We will mostly follow the notations
and conventions used there.

The ten-dimensional spacetime is $\mathbb{S}_{t}^{1}\times\mathbb{C}^{4}\times\mathbb{R}$.
The coordinates on $\mathbb{S}_{t}^{1}$ and $\mathbb{R}$ are taken
to be $x^{0}$ and $x^{9}$, respectively. We denote the set of coordinate
labels of four complex planes by 
\begin{equation}
\underline{4}=\left\{ 1,2,3,4\right\} ,
\end{equation}
and the complex coordinate on $\mathbb{C}_{a}\subset\mathbb{C}^{4}$
by 
\begin{equation}
z_{a}=x^{2a-1}+\mathrm{i}x^{2a}.
\end{equation}
For each 
\begin{equation}
A\in\underline{4}^{\vee}=\left\{ \left(123\right),\left(124\right),\left(134\right),\left(234\right)\right\} ,
\end{equation}
we define the complex three-plane 
\begin{equation}
\mathbb{C}_{A}^{3}=\prod_{a\in A}\mathbb{C}_{a}.
\end{equation}
The complex plane that is complementary to $\mathbb{C}_{A}^{3}$ is
$\mathbb{C}_{\check{A}}$ with 
\begin{equation}
\check{A}=\left\{ \left.a\in\underline{4}\right|a\notin A\right\} .
\end{equation}

We introduce $k$ D$0$-branes along $\mathbb{S}_{t}^{1}$ and $n_{A}$
D$6_{A}$-branes along $\mathbb{S}_{t}^{1}\times\mathbb{C}_{A}^{3}$
for all $A\in\underline{4}^{\vee}$. We will use the notation
\begin{equation}
\vec{n}=\left(n_{123},n_{124},n_{134,}n_{234}\right)
\end{equation}
to capture the number of D$6$-branes of different stacks. In total,
there are four stacks of intersecting D$6$-branes. A summary of the
configuration is in Table \ref{D0-D6}. 

The presence of the D$0$-branes and the D$6$-branes breaks the ten-dimensional
Lorentz group $\mathrm{SO}\left(1,9\right)$ down to $\prod_{a\in\underline{4}}\mathrm{U}\left(1\right)_{a}$,
where $\mathrm{U}\left(1\right)_{a}$ rotates the complex plane $\mathbb{C}_{a}$.
We turn on a constant Neveu-Schwarz $B$-field along $\mathbb{C}^{4}$
\cite{Seiberg:1999vs,Witten:2000mf},
\begin{equation}
B=\sum_{a\in\underline{4}}b_{a}dx^{2a-1}\wedge dx^{2a},\quad b_{a}=-\cot\vartheta_{a},\quad0<\vartheta_{a}<\pi.\label{eq:BTetrahedron}
\end{equation}
As analyzed in \cite{Pomoni:2021hkn}, when
\begin{equation}
\frac{2\pi}{3}<\vartheta_{1}=\vartheta_{2}=\vartheta_{3}=\vartheta_{4}<\pi,\label{eq:StrongB}
\end{equation}
the original string theory vacuum is unstable and supersymmetry is
completely broken, but the effect of tachyon condensation restores
a part of supersymmetry. After rolling to the true vacuum, there are
two unbroken supercharges. 

We can have two different ways to understand this setup. The first
one is the viewpoint of conventional field theory. Without loss of
generality, we can take the physical spacetime to be $\mathbb{S}_{t}^{1}\times\mathbb{C}_{123}^{3}$.
The bound states of D$0$-branes and D$6_{123}$-branes realize noncommutative
instantons in the $(6+1)$-dimensional super Yang-Mills theory, while
the effect of the other D$6$-branes is to produce defects of real
codimension two. Alternatively, we can adopt the interpretation of
generalized field theory, which can be constructed by merging several
ordinary field theories across defects \cite{Nekrasov:2016qym,Nekrasov:2016gud,Nekrasov:2017rqy}.
The spacetime of a generalized field theory is generally not a manifold,
but a union of several intersecting components. In our case, the spacetime
is
\begin{equation}
\bigcup_{A\in\underline{4}^{\vee}}\mathbb{C}_{A}^{3}\subset\mathbb{C}^{4}.
\end{equation}
The fields and the gauge groups on different components can be different,
and the matter fields living on the intersection $\mathbb{C}_{A}^{3}\cap\mathbb{C}_{B}^{3}$
for $A\neq B$ transform in the bifundamental representation of the
product group $\mathrm{U}\left(n_{A}\right)\times\mathrm{U}\left(n_{B}\right)$.
In \cite{Pomoni:2021hkn}, we described this generalized field theory
in the framework of noncommutative field theory. 

\begin{table}
\begin{centering}
\begin{tabular}{|c|c|c|c|}
\hline 
Strings & $\mathcal{N}=4$ & $\mathcal{N}=2$ & $\left(\mathrm{U}\left(k\right),\mathrm{U}\left(n_{A}\right)\right)$\tabularnewline
\hline 
\multirow{4}{*}{D$0$-D$0$} & \multirow{2}{*}{Vector} & Vector $\varUpsilon$ & \multirow{4}{*}{$\left(\mathrm{\mathbf{Adj}},\mathbf{1}\right)$}\tabularnewline
\cline{3-3} 
 &  & Chiral $\varPhi_{\breve{A}}=B_{\breve{A}}+\cdots$ & \tabularnewline
\cline{2-3} \cline{3-3} 
 & \multirow{2}{*}{Chiral $\Phi_{a}$ $(a\in A)$} & Chiral $\varPhi_{a}=B_{a}+\cdots$ & \tabularnewline
\cline{3-3} 
 &  & Fermi $\varPsi_{a}=\psi_{a}+\cdots$ & \tabularnewline
\hline 
\multirow{2}{*}{D$0$-D$6_{A}$} & \multirow{2}{*}{Chiral $\Phi_{A}$} & Chiral $\varPhi_{A}=I_{A}+\cdots$ & \multirow{2}{*}{$\left(\mathbf{k},\overline{\mathbf{n}_{A}}\right)$}\tabularnewline
\cline{3-3} 
 &  & Fermi $\varPsi_{A}=\psi_{A}+\cdots$ & \tabularnewline
\hline 
\end{tabular}
\par\end{centering}
\caption{Field content from the quantization of D$0$-D$0$ and D$0$-D$6_{A}$
strings in terms of $\mathcal{N}=4$ and $\mathcal{N}=2$ multiplets.
The $\mathcal{N}=2$ chiral and Fermi superfields satisfy $\bar{\mathcal{D}}\varPhi_{\breve{A}}=\bar{\mathcal{D}}\varPhi_{a}=\bar{\mathcal{D}}\varPhi_{A}=0$,
$\bar{\mathcal{D}}\varPsi_{a}=\sqrt{2}E_{a}=\sqrt{2}\left[\varPhi_{\breve{A}},\varPhi_{a}\right]$,
and $\bar{\mathcal{D}}\varPsi_{A}=\sqrt{2}E_{A}=\sqrt{2}\varPhi_{\breve{A}}\varPhi_{A}$.
Here $\mathcal{D}$ and $\bar{\mathcal{D}}$ are the gauge-covariant
super-derivatives. \label{multiplets}}
\end{table}

\begin{figure}
\centering
\includegraphics[width=0.3\textwidth]{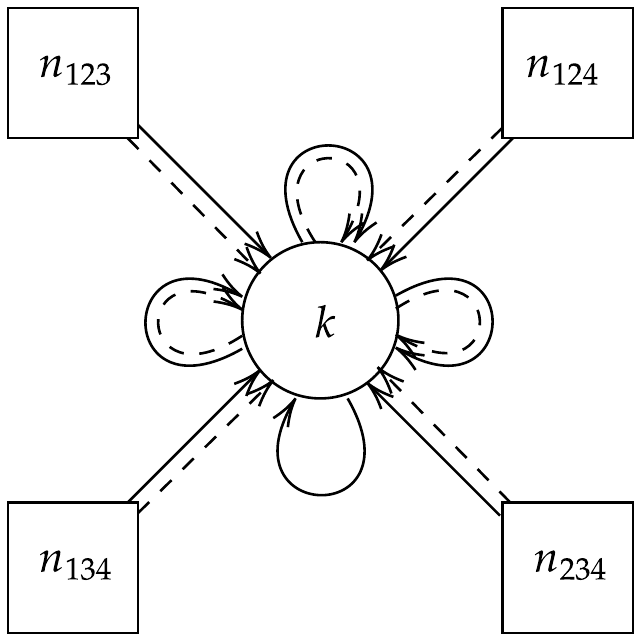}

\caption{The quiver diagram of the low-energy theory $\mathcal{T}_{\vec{n},k}$
on $k$ D$0$-branes probing $n_{A}$ D$6_{A}$-branes for $A\in\underline{4}^{\vee}$.
Each solid line represents an $\mathcal{N}=2$ chiral superfield,
while each dashed line represents an $\mathcal{N}=2$ Fermi superfield.}
\label{quiverdiagram}
\end{figure}

We are interested in the low-energy effective theory $\mathcal{T}_{\vec{n},k}$
on the worldline of D$0$-branes. The field content is summarized
in Table \ref{multiplets}. The quantization of D$0$-D$0$ strings
gives rise to an $\mathcal{N}=4$ $\mathrm{U}(k)$ vector multiplet
and three massless $\mathcal{N}=4$ chiral multiplets $\Phi_{a},a\in A$
in the adjoint representation of $\mathrm{U}(k)$. There is a superpotential
\begin{equation}
\mathcal{W}=\frac{1}{6}\epsilon^{abc}\mathrm{Tr}\left(\Phi_{a}\left[\Phi_{b},\Phi_{c}\right]\right),\quad A=(abc),
\end{equation}
which arises from the dimensional reduction of the ten-dimensional
$\mathrm{U}\left(k\right)$ super Yang-Mills theory. The quantization
of D$0$-D$6_{A}$ and D$6_{A}$-D$0$ strings leads to a massless
$\mathcal{N}=4$ chiral multiplet $\Phi_{A}$, transforming in the
bifundamental representation of $\mathrm{U}\left(k\right)\times\mathrm{U}\left(n_{A}\right)$.
Here the $\mathrm{U}\left(n_{A}\right)$ symmetry, which is supported
by the D$6_{A}$-branes, appears as a global symmetry from the perspective
of D$0$-brane quantum mechanics. Notice that the $\mathcal{N}=4$
supersymmetry preserved by the D$0$-branes and the D$6_{A}$-branes
is different from that by the D$0$-branes and the D$6_{B}$-branes
for $A\neq B\in\underline{4}^{\vee}$, and only an $\mathcal{N}=2$
supersymmetry is shared \cite{Pomoni:2021hkn}. Hence, $\mathcal{T}_{\vec{n},k}$
is an $\mathcal{N}=2$ supersymmetric quantum mechanics. In the language
of $\mathcal{N}=2$ superspace, the Lagrangian is given by
\begin{align}
\mathcal{L}= & \int d\theta d\bar{\theta}\mathrm{Tr}\left(\frac{1}{2e^{2}}\bar{\varUpsilon}\varUpsilon-\frac{\mathrm{i}}{4}\sum_{a\in\underline{4}}\bar{\varPhi}_{a}\mathcal{D}_{t}\varPhi_{a}-\frac{1}{2}\sum_{i=1}^{3}\bar{\varPsi}_{i}\varPsi_{i}\right)\nonumber \\
 & -\frac{1}{2\sqrt{2}}\mathrm{Tr}\left(\int d\theta\left.\sum_{i,j,k=1}^{3}\epsilon^{ijk}\varPsi_{i}\left[\varPhi_{j},\varPhi_{k}\right]\right|_{\bar{\theta}=0}+c.c.\right)\nonumber \\
 & +\left(\frac{\mathrm{i}r}{2}\int d\theta\left.\varUpsilon\right|_{\bar{\theta}=0}+\mathrm{c.c.}\right)\nonumber \\
 & -\mathrm{Tr}\sum_{A\in\underline{4}^{\vee}}\left(\frac{\mathrm{i}}{4}\bar{\varPhi}_{A}\mathcal{D}_{t}\varPhi_{A}+\frac{1}{2}\bar{\varPsi}_{A}\varPsi_{A}\right),\label{eq:Lnk}
\end{align}
where $\mathcal{D}_{t}$ is the covariant time derivative, and the
parameter $r$ is positive when the condition (\ref{eq:StrongB})
is satisfied. The corresponding quiver diagram is presented in Figure
\ref{quiverdiagram}. After integrating out the auxiliary fields,
we can obtain the scalar potential 
\begin{align}
V= & \mathrm{Tr}\left(\sum_{a\in\underline{4}}\left[B_{a},B_{a}^{\dagger}\right]+\sum_{A\in\underline{4}^{\vee}}I_{A}I_{A}^{\dagger}-r\cdot\mathbb{I}\right)^{2}\nonumber \\
 & +\sum_{A\in\underline{4}^{\vee}}\mathrm{Tr}\left|B_{\breve{A}}I_{A}\right|^{2}+\sum_{a<b\in\underline{4}}\mathrm{Tr}\left|\left[B_{a},B_{b}\right]\right|^{2},
\end{align}
which leads to the moduli space $\mathfrak{M}_{\vec{n},k}$ of the
supersymmetric ground states
\begin{align}
\mathfrak{M}_{\vec{n},k}= & \left.\left\{ \left.\left(\vec{B},\vec{I}\right)\right|V=0\right\} \right/\mathrm{U}(k)\nonumber \\
= & \left.\left\{ \left.\left(\vec{B},\vec{I}\right)\right|\mu^{\mathbb{R}}-r\cdot\mathbb{I}=\mu^{\mathbb{C}}=\sigma=0\right\} \right/\mathrm{U}(k),\label{eq:moduli}
\end{align}
where
\begin{align}
\vec{B}= & \left(B_{a}\right)_{a\in\underline{4}},\quad\vec{I}=\left(I_{A}\right)_{A\in\underline{4}^{\vee}},\\
\mu^{\mathbb{R}}= & \sum_{a\in\underline{4}}\left[B_{a},B_{a}^{\dagger}\right]+\sum_{A\in\underline{4}^{\vee}}I_{A}I_{A}^{\dagger},\\
\mu^{\mathbb{C}}= & \left(\mu_{ab}^{\mathbb{C}}=\left[B_{a},B_{b}\right]\right)_{a,b\in\underline{4}},\\
\sigma= & \left(\sigma_{A}=B_{\check{A}}I_{A}\right)_{A\in\underline{4}^{\vee}},
\end{align}
and the $\mathrm{U}(k)$ symmetry acts on $B_{a}$ in the adjoint
representation and $I_{A}$ in the fundamental representation, 
\begin{equation}
\left(B_{a},I_{A}\right)\to\left(gB_{a}g^{-1},gI_{A}\right),\quad g\in\mathrm{U}(k).
\end{equation}
In \cite{Pomoni:2021hkn}, $\mathfrak{M}_{\vec{n},k}$ is called the
moduli space of tetrahedron instantons with instanton number $k$.

When $\vec{n}=\left(n,0,0,0\right)$, all D$6$-branes are parallel,
and the system preserves four instead of two supercharges. It provides
a string theory realization of K-theoretic Donaldson-Thomas theory
on $\mathbb{C}^{3}$ \cite{maulik2006gromov1,maulik2006gromov2,Cirafici:2008sn}.
The moduli space $\mathfrak{M}_{\left(n,0,0,0\right),k}$ coincides
with the  moduli space of torsion free sheaves $\mathcal{E}$ on $\mathbb{CP}^{3}=\mathbb{C}^{3}\cup\mathbb{CP}_{\infty}^{2}$
with the Chern character $\mathrm{ch}\left(\mathcal{E}\right)=\left(n,0,0,-k\right)$
and the framing $\left.\Phi\right|_{\mathbb{CP}_{\infty}^{2}}\cong\mathbb{C}^{n}\otimes\mathcal{O}_{\mathbb{CP}_{\infty}^{2}}$.
This moduli space is also isomorphic to a Quot scheme $\mathrm{Quot}_{\mathbb{C}^{3}}^{k}\left(\mathcal{O}^{\oplus n}\right)$,
which reduces to the Hilbert scheme $\mathrm{Hilb}^{k}\left(\mathbb{C}^{3}\right)$
of $k$ points on $\mathbb{C}^{3}$ when $n=1$ \cite{Cazzaniga:2020xru}.
For the general case of $\mathfrak{M}_{\vec{n},k}$, we sketched a
geometric interpretation in the framework of Quot schemes in \cite{Pomoni:2021hkn}.
A careful mathematical treatment was given recently in \cite{Fasola:2023ypx}.

\section{Tetrahedron instanton partition function \label{sec:PartitionFunction}}

The D$0$-D$6$ partition function in the $\Omega$-background is
computed by the refined Witten index of the supersymmetric quantum
mechanics $\mathcal{T}_{\vec{n},k}$, 
\begin{equation}
\mathcal{Z}_{\vec{n},k}=\mathrm{Tr}_{\mathcal{H}_{\vec{n},k}}\left(-1\right)^{F}e^{-\beta\left\{ \mathcal{Q},\bar{\mathcal{Q}}\right\} }\left(\prod_{A\in\underline{4}^{\vee}}\prod_{\alpha=1}^{n_{A}}t_{A,\alpha}^{T_{A,\alpha}}\right)\left(\prod_{a\in\underline{4}}q_{a}^{\mathcal{J}_{a}}\right),\label{eq:Znk}
\end{equation}
with
\begin{equation}
\prod_{a\in\underline{4}}q_{a}=1.\label{eq:prodqa}
\end{equation}
Here $\mathcal{H}_{\vec{n},k}$ is the Hilbert space of $\mathcal{T}_{\vec{n},k}$
on $\mathbb{S}_{t}^{1}$, $F$ is the fermion number operator, $\beta$
is the circumference of $\mathbb{S}_{t}^{1}$, $\mathcal{Q}$ and
$\bar{\mathcal{Q}}$ are two conserved supercharges, $T_{A,\alpha},\alpha=1,\cdots,n_{A}$
are the Cartan generators of $\mathrm{U}\left(n_{A}\right)$, and
$\mathcal{J}_{a}$ is the generator of the $\mathrm{U}\left(1\right)_{a}$
symmetry of $\mathcal{T}_{\vec{n},k}$ originating from the symmetry
rotating $\mathbb{C}_{a}$. The supercharges $\mathcal{Q}$ and $\bar{\mathcal{Q}}$
commute with $T_{A,\alpha}$, but not with $\mathcal{J}_{a}$,
\begin{equation}
\left[\mathcal{J}_{a},\mathcal{Q}\right]=-\frac{1}{2}\mathcal{Q},\quad\left[\mathcal{J}_{a},\bar{\mathcal{Q}}\right]=\frac{1}{2}\bar{\mathcal{Q}}.\label{eq:JQ}
\end{equation}
Thus, we can at most have three linearly independent combinations
of $\mathcal{J}_{a}$ that are $\mathcal{Q}$-closed. One possible
choice is $\left\{ \mathcal{J}_{1}-\mathcal{J}_{4},\mathcal{J}_{2}-\mathcal{J}_{4},\mathcal{J}_{3}-\mathcal{J}_{4}\right\} $.
Equivalently, we can keep all $\mathcal{J}_{a},a\in\underline{4}$
and impose the condition (\ref{eq:prodqa}) on $q_{a}$. Many interesting
geometric properties of $\mathfrak{M}_{\vec{n},k}$ are captured by
$\mathcal{Z}_{\vec{n},k}$. Applying the supersymmetric localization
techniques, we can write $\mathcal{Z}_{\vec{n},k}$ as a contour integral
\cite{Pomoni:2021hkn}, 
\begin{align}
\mathcal{Z}_{\vec{n},k}= & \frac{1}{k!}\oint_{\mathrm{JK}}\prod_{i=1}^{k}\frac{du_{i}}{2\pi\mathrm{i}u_{i}}\left(\prod_{\substack{i,j=1\\
i\neq j
}
}^{k}\left\llbracket u_{i}u_{j}^{-1}\right\rrbracket \right)\nonumber \\
 & \times\left(\prod_{i,j=1}^{k}\frac{\prod_{1\leq a<b\leq3}\left\llbracket q_{a}q_{b}u_{i}u_{j}^{-1}\right\rrbracket }{\prod_{a\in\underline{4}}\left\llbracket q_{a}u_{i}u_{j}^{-1}\right\rrbracket }\right)\left(\prod_{i=1}^{k}\prod_{A\in\underline{4}^{\vee}}\prod_{\alpha=1}^{n_{A}}\frac{\left\llbracket q_{\check{A}}u_{i}t_{A,\alpha}^{-1}\right\rrbracket }{\left\llbracket u_{i}t_{A,\alpha}^{-1}\right\rrbracket }\right),\label{eq:ZnkContour}
\end{align}
where the contour is specified by the Jeffrey-Kirwan residue prescription
\cite{jeffrey1995localization}, and 
\begin{equation}
\left\llbracket x\right\rrbracket =x^{\frac{1}{2}}-x^{-\frac{1}{2}}=-\left\llbracket x^{-1}\right\rrbracket .
\end{equation}
The genuine poles of the contour integral are classified by a collection
of colored plane partitions
\begin{equation}
\vec{\pi}=\left\{ \pi^{\left(A,\alpha\right)},A\in\underline{4}^{\vee},\alpha=1,\cdots,n_{A},\left|\vec{\pi}\right|=k\right\} ,
\end{equation}
where each $\pi^{\left(A,\alpha\right)}$ is a plane partition, and
$\left|\vec{\pi}\right|$ is the total number of boxes of $\vec{\pi}$.
Every plane partition $\pi^{\left(A,\alpha\right)}$ can be visualized
as a set of boxes sitting in $\mathbb{Z}_{+}^{3}$,
\begin{equation}
\pi^{\left(A,\alpha\right)}=\left\{ \left.\left(x,y,z\right)\in\mathbb{Z}_{+}^{3}\right|1\leq z\leq\pi^{\left(A,\alpha\right)}\right\} ,
\end{equation}
so that there can be at most one box at $\left(x,y,z\right)$, and
a box can occupy $\left(x,y,z\right)$ only if there are boxes in
$\left(x^{\prime},y,z\right),\left(x,y^{\prime},z\right),\left(x,y,z^{\prime}\right)$
for all $1\leq x^{\prime}<x$, $1\leq y^{\prime}<y$, $1\leq z^{\prime}<z$. 

The K-theoretic tetrahedron instanton partition function $\mathcal{Z}_{\vec{n}}$
is defined by packaging $\mathcal{Z}_{\vec{n},k}$ into a generating
function,
\begin{equation}
\mathcal{Z}_{\vec{n}}=1+\sum_{k=1}^{\infty}\left(-p\right)^{k}\mathcal{Z}_{\vec{n},k},\label{eq:Zinst}
\end{equation}
where the instanton counting parameter is taken to be $-p$ for convenience.
In terms of the characters
\begin{align}
\mathbf{N}_{A} & =\sum_{\alpha=1}^{n_{A}}t_{A,\alpha},\\
\mathbf{K}_{\vec{\pi}} & =\sum_{A=(a,b,c)\in\underline{4}^{\vee}}\sum_{\alpha=1}^{n_{A}}\sum_{(x,y,z)\in\pi^{(A,\alpha)}}t_{A,\alpha}q_{a}^{1-x}q_{b}^{1-y}q_{c}^{1-z},\\
\mathbf{L} & =1-\sum_{a\in\underline{4}}q_{a}^{-1}+\sum_{1\leq a<b\leq3}q_{a}^{-1}q_{b}^{-1},
\end{align}
we can express $\mathcal{Z}_{\vec{n}}$ as an infinite sum over a
collection of colored plane partitions \cite{Pomoni:2021hkn},
\begin{equation}
\mathcal{Z}_{\vec{n}}=\sum_{\vec{\pi}}\left(-p\right)^{\left|\vec{\pi}\right|}\mathtt{PE}\left[-\mathbf{K}_{\vec{\pi}}^{\vee}\mathbf{K}_{\vec{\pi}}\mathbf{L}+\sum_{A\in\underline{4}^{\vee}}\mathbf{N}_{A}^{\vee}\mathbf{K}_{\vec{\pi}}\left(1-q_{A}^{-1}\right)-\mathrm{constant}\right],\label{eq:Znsum}
\end{equation}
where the constant part that is independent of $t_{A,\alpha}$ and
$q_{a}$ is subtracted. The plethystic exponential operator $\mathtt{PE}$
is defined by
\begin{equation}
\mathtt{PE}\left[f\left(x_{1},\cdots,x_{n}\right)\right]=\exp\left(\sum_{\ell=1}^{\infty}\frac{1}{\ell}f\left(x_{1}^{\ell},\cdots,x_{n}^{\ell}\right)\right),
\end{equation}
and the dual operator $^{\vee}$ acts on a character by
\begin{equation}
\left(\sum_{a}x_{a}\right)^{\vee}=\sum_{a}x_{a}^{-1}.
\end{equation}

We would like to point out that we define $\mathcal{Z}_{\vec{n},k}$
in such a way that makes formulas simple and puts four stacks of D$6$-branes
on equal footing. This definition can be different from those used
in \cite{Cirafici:2008sn,Awata:2009dd,Fasola:2023ypx} by a sign.
However, the difference can be reconciled by a redefinition of $p$.

The expression (\ref{eq:Znsum}) looks quite intricate, and appears
to depend on all the parameters in the definition (\ref{eq:Znk}).
In particular, each term in (\ref{eq:Znsum}) depends on all the Coulomb
parameters $t_{A,\alpha}$ except for an overall factor associated
with the center $\mathrm{U}(1)_{c}$ of $\prod_{A\in\underline{4}^{\vee}}\mathrm{U}\left(n_{A}\right)$.
However, we find that the dependence on all the Coulomb parameters
$t_{A,\alpha}$ disappears if we combine the contributions associated
with all $\vec{\pi}$ with fixed $\left|\vec{\pi}\right|$. This means
that $\mathcal{Z}_{\vec{n}}$ does not depend on the positions of
D$6$-branes at all. Inspired by \cite{nekrasov2009instanton,Nekrasov:2014nea,Benini:2018hjy,Nekrasov:2017cih,Nekrasov:2018xsb},
we propose that (\ref{eq:Znsum}) has a simple closed-form expression
\begin{equation}
\mathcal{Z}_{\vec{n}}=\mathtt{PE}\left[\mathcal{F}_{\vec{n}}\left(q_{1},q_{2},q_{3},q_{4},p\right)\right],\label{eq:ZnPE}
\end{equation}
where 
\begin{equation}
\boxed{\mathcal{F}_{\vec{n}}\left(q_{1},q_{2},q_{3},q_{4},p\right)=\frac{\prod_{1\leq a<b\leq3}\left\llbracket q_{a}q_{b}\right\rrbracket }{\prod_{a\in\underline{4}}\left\llbracket q_{a}\right\rrbracket }\frac{\left\llbracket \varDelta_{\vec{n}}\right\rrbracket }{\left\llbracket \varDelta_{\vec{n}}^{\frac{1}{2}}p\right\rrbracket \left\llbracket \varDelta_{\vec{n}}^{\frac{1}{2}}p^{-1}\right\rrbracket }}.\label{eq:Fgeneral}
\end{equation}
Here $\varDelta_{\vec{n}}$ depends on $\vec{n}$ and the $\Omega$-deformation
parameters,
\begin{equation}
\varDelta_{\vec{n}}=\prod_{A\in\underline{4}^{\vee}}q_{\check{A}}^{n_{A}}.\label{eq:Deltan}
\end{equation}
We have checked (\ref{eq:ZnPE}) up to $k=4$ for various $\vec{n}$
with $n_{A}\leq3$. We believe that a rigorous proof of (\ref{eq:ZnPE})
can be given by extending the derivation in \cite{Fasola:2020hqa}.
We will not explore it in this paper, but it would be interesting
to do so. \footnote{Shortly after submitting this paper, this closed-form expression was
proved by Fasola and Monavari in \cite{Fasola:2023ypx}.}

When $\vec{n}=\left(n,0,0,0\right)$, we can recover the previous
result obtained in \cite{nekrasov2009instanton,Nekrasov:2014nea,Awata:2009dd},
\begin{align}
\mathcal{Z}_{n} & =\mathcal{Z}_{\left(n,0,0,0\right)}\nonumber \\
 & =\sum_{\vec{\pi}}\left(-p\right)^{\left|\vec{\pi}\right|}\mathtt{PE}\left[\mathbf{N}_{123}^{\vee}\mathbf{K}_{\vec{\pi}}-q_{1}q_{2}q_{3}\mathbf{N}_{123}\mathbf{K}_{\vec{\pi}}^{\vee}-\prod_{a=1}^{3}\left(1-q_{a}\right)\mathbf{K}_{\vec{\pi}}^{\vee}\mathbf{K}_{\vec{\pi}}-\mathrm{constant}\right]\nonumber \\
 & =\mathtt{PE}\left[\mathcal{F}_{n}\left(q_{1},q_{2},q_{3},q_{4},p\right)\right],
\end{align}
where
\begin{equation}
\mathcal{F}_{n}\left(q_{1},q_{2},q_{3},q_{4},p\right)=\frac{\left\llbracket q_{1}q_{2}\right\rrbracket \left\llbracket q_{1}q_{3}\right\rrbracket \left\llbracket q_{2}q_{3}\right\rrbracket }{\left\llbracket q_{1}\right\rrbracket \left\llbracket q_{2}\right\rrbracket \left\llbracket q_{3}\right\rrbracket \left\llbracket q_{4}^{\frac{n}{2}}p\right\rrbracket \left\llbracket q_{4}^{\frac{n}{2}}p^{-1}\right\rrbracket }\frac{\left\llbracket q_{4}^{n}\right\rrbracket }{\left\llbracket q_{4}\right\rrbracket },\label{eq:Fn}
\end{equation}
which further reduces to the case when $n=1$ as
\begin{equation}
\mathcal{F}_{1}\left(q_{1},q_{2},q_{3},q_{4},p\right)=\frac{\left\llbracket q_{1}q_{2}\right\rrbracket \left\llbracket q_{1}q_{3}\right\rrbracket \left\llbracket q_{2}q_{3}\right\rrbracket }{\left\llbracket q_{1}\right\rrbracket \left\llbracket q_{2}\right\rrbracket \left\llbracket q_{3}\right\rrbracket \left\llbracket q_{4}^{\frac{1}{2}}p\right\rrbracket \left\llbracket q_{4}^{\frac{1}{2}}p^{-1}\right\rrbracket }.\label{eq:F1}
\end{equation}
We would like to emphasize that $\mathcal{F}_{n}$ has two classes
of singularities: $q_{a}\to1,a=1,2,3$ and $p\to q_{4}^{\pm\frac{n}{2}}$.
The first class is simply the standard thermodynamic limit in which
D$0$-branes are allowed to move freely along the $a$th complex plane
rather than being confined to its origin by the $\Omega$-deformation.
The second class is more interesting, and the existence of such singularities
implies that extra flat directions appear in the finite coupling regime.
The limit $q_{4}\to1$ is smooth, and is the Calabi-Yau condition
for $\mathbb{C}_{123}$. Thus there are five flat complex directions
in total. An explanation for the singularity structure of $\mathcal{F}_{n}$
can be given from the M-theory perspective.

\section{Connection with the magnificent four model \label{sec:MF}}

The simple formula (\ref{eq:Fgeneral}) is indicative of a connection
with a closely related model, the so-called magnificent four model
\cite{Nekrasov:2017cih}, which can be constructed by a system of
$k$ D$0$-branes probing $N$ D$8$-branes and $N$ anti-D$8$-branes
on $\mathbb{S}_{t}^{1}\times\mathbb{C}^{4}\times\mathbb{R}$. This
configuration preserves two supercharges when a suitable constant
$B$-field is turned on \cite{Witten:2000mf}. The low-energy effective
theory is described by an $\mathcal{N}=2$ supersymmetric quantum
mechanics $\mathcal{T}_{k}^{\mathrm{MF}}$. The partition function
$\mathcal{Z}^{\mathrm{MF}}$ of the magnificent four model is the
generating function of the refined Witten index $\mathcal{Z}_{k}^{\mathrm{MF}}$
of $\mathcal{T}_{k}^{\mathrm{MF}}$,
\begin{equation}
\mathcal{Z}^{\mathrm{MF}}\left(q_{1},q_{2},q_{3},q_{4},p\right)=1+\sum_{k=1}^{\infty}\left(-p\right)^{k}\mathcal{Z}_{k}^{\mathrm{MF}}.
\end{equation}
Here 
\begin{equation}
\mathcal{Z}_{k}^{\mathrm{MF}}=\mathrm{Tr}_{\mathcal{H}_{k}^{\mathrm{MF}}}\left(-1\right)^{F}g\left(\prod_{a=1}^{4}q_{a}^{\mathcal{J}_{a}}\right),\quad\prod_{a=1}^{4}q_{a}=1.
\end{equation}
where $\mathcal{H}_{k}^{\mathrm{MF}}$ is the Hilbert space of $\mathcal{T}_{k}^{\mathrm{MF}}$
on $\mathbb{S}_{t}^{1}$, $g$ is an element of the global symmetry
group $\mathrm{U}\left(N\right)\times\mathrm{U}\left(N\right)\subset\mathrm{U}\left(\left.N\right|N\right)$
associated with D$8$-branes and anti-D$8$-branes, $\mathcal{J}_{a}$
rotates the $a$th complex plane $\mathbb{C}_{a}\subset\mathbb{C}^{4}$,
and $\left(q_{1},q_{2},q_{3},q_{4}\right)\in\mathrm{U}\left(1\right)^{3}\subset\mathrm{SU}\left(4\right)$
are the $\Omega$-deformation parameters. The partition function $\mathcal{Z}^{\mathrm{MF}}$
also admits a closed-form expression \cite{Nekrasov:2017cih,Nekrasov:2018xsb},
\begin{equation}
\mathcal{Z}^{\mathrm{MF}}\left(q_{1},q_{2},q_{3},q_{4},\mu,p\right)=\mathtt{PE}\left[\frac{\prod_{1\leq a<b\leq3}\left\llbracket q_{a}q_{b}\right\rrbracket }{\prod_{a\in\underline{4}}\left\llbracket q_{a}\right\rrbracket }\frac{\left\llbracket s\right\rrbracket }{\left\llbracket s^{\frac{1}{2}}p\right\rrbracket \left\llbracket s^{\frac{1}{2}}p^{-1}\right\rrbracket }\right],\label{eq:ZMF}
\end{equation}
where $s$ depends only on Coulomb branch parameters associated with
$\mathrm{U}\left(N\right)\times\mathrm{U}\left(N\right)$.

Comparing (\ref{eq:ZMF}) and (\ref{eq:Fgeneral}), we immediately
find that
\begin{equation}
\mathcal{Z}^{\mathrm{MF}}\left(q_{1},q_{2},q_{3},q_{4},s=\varDelta_{\vec{n}},p\right)=\mathcal{Z}_{\vec{n}}\left(q_{1},q_{2},q_{3},q_{4},p\right),\label{eq:ZMFZn}
\end{equation}
which generalizes the known agreement for $\vec{n}=\left(n,0,0,0\right)$
\cite{Nekrasov:2018xsb}. This suggests that the annihilation of D$8$-branes
and anti-D$8$-branes in the $\Omega$-background happens not only
when they coincide but also when they are separated by a suitable
distance. Furthermore, a system of D$6$-branes will be produced after
the annihilation. It is interesting that the configuration of D$6$-branes
can be either parallel or intersecting, depending on the distance
of the D$8$-branes and the anti-D$8$-branes when the annihilation
happens.

\section{The index of M-theory \label{sec:M-theory}}

In the previous sections, we take the perspective of type IIA superstring
theory. In this section, we shift our attention to M-theory and introduce
the index of M-theory on $\mathscr{X}$, where $\mathscr{X}$ is a
noncompact toric Calabi-Yau fivefold.

The eleven-dimensional Minkowski vacuum of M-theory preserves an eleven-dimensional
Majorana spinor supercharge with $32$ real components, among which
only a part is preserved when M-theory lives in $\mathbb{S}_{t}^{1}\times\mathscr{X}$.
According to the branching rule for $\mathrm{Spin}(1,10)\supset\mathrm{Spin}(10)\supset\mathrm{SU}(5)\times\mathrm{U}(1)_{R}$,
\begin{align}
\mathbf{32}_{\mathrm{Spin}(1,10)} & \to\left[\mathbf{16}+\bar{\mathbf{16}}\right]_{\mathrm{Spin}(10)}\nonumber \\
 & \to\left[\left(\mathbf{1}_{-5}+\bar{\mathbf{5}}_{3}+\mathbf{10}_{-1}\right)+\left(\mathbf{1}_{5}+\mathbf{5}_{-3}+\bar{\mathbf{10}}_{1}\right)\right]_{\mathrm{SU}(5)\times\mathrm{U}(1)_{R}},
\end{align}
the eleven-dimensional supercharge can be decomposed as 
\begin{equation}
\left(\mathscr{Q}\oplus\mathscr{Q}_{I}\oplus\mathscr{Q}^{[IJ]}\right)\oplus\left(\bar{\mathscr{Q}}\oplus\bar{\mathscr{Q}}^{I}\oplus\bar{\mathscr{Q}}_{[IJ]}\right),\quad I,J=1,\cdots,5.
\end{equation}
In general, only two supercharges $\mathscr{Q}$ and $\bar{\mathscr{Q}}$
associated with representations $\mathbf{1}_{\pm5}$ are unbroken.
Meanwhile, from the branching rule
\begin{align}
\mathbf{11}_{\mathrm{Spin}(1,10)} & \to\left[\mathbf{10}+\mathbf{1}\right]_{\mathrm{Spin}(10)}\nonumber \\
 & \to\left[\left(\mathbf{5}_{2}+\bar{\mathbf{5}}_{-2}\right)+\mathbf{1}_{0}\right]_{\mathrm{SU}(5)\times\mathrm{U}(1)_{R}},
\end{align}
only the time translation component $\mathscr{H}=-\mathrm{i}\partial_{t}$
of the eleven-dimensional momentum remains. It follows from the eleven-dimensional
supersymmetry algebra that the effective one-dimensional theory on
$\mathbb{S}_{t}^{1}$ enjoys an $\mathcal{N}=2$ supersymmetry algebra,
\begin{equation}
\left\{ \mathscr{Q},\bar{\mathscr{Q}}\right\} =\mathscr{H},
\end{equation}
and $\mathrm{U}(1)_{R}$ is identified with the R-symmetry. We would
obtain a local $\mathcal{N}=2$ supersymmetry in one dimension if
$\mathscr{X}$ was a compact Calabi-Yau fivefold. Although the component
fields of the one-dimensional gravity multiplet are not dynamical,
they generate constraints and cannot be neglected \cite{Haupt:2008nu}.
On the contrary, since $\mathscr{X}$ of interest are noncompact,
the gravitational effects are turned off, and the low-energy dynamics
is captured by an $\mathcal{N}=2$ supersymmetric quantum mechanics
$\mathbf{SQM}\left(\mathscr{X}\right)$ on $\mathbb{S}_{t}^{1}$. 

Furthermore, since $\mathscr{X}$ is a toric manifold, the $\mathrm{U}(1)^{5}$
isometries lead to extra global symmetries in $\mathbf{SQM}\left(\mathscr{X}\right)$
with generators $\mathscr{J}_{I},I=1,\cdots,5$. None of $\mathscr{J}_{I}$
commutes with $\mathscr{Q}$ and $\bar{\mathscr{Q}}$, but there are
four linearly independent combinations of $\mathscr{J}_{I}$ that
commute with $\mathscr{Q}$ and $\bar{\mathscr{Q}}$. This is consistent
with the fact that only $\mathrm{U}(1)^{4}\subset\mathrm{U}(1)^{5}$
preserves the nowhere-vanishing holomorphic top-form $\varOmega$
of $\mathscr{X}$. 

We define the index of M-theory on $\mathscr{X}$ to be the refined
Witten index of $\mathbf{SQM}\left(\mathscr{X}\right)$,
\begin{equation}
\mathscr{Z}_{\mathscr{X}}\left(v_{1},v_{2},v_{3},v_{4},v_{5}\right)=\mathrm{Tr}_{\mathcal{H}(\mathscr{X})}(-1)^{F}e^{-\beta\mathscr{H}}\left(\prod_{I=1}^{5}v_{I}^{\mathscr{J}_{I}}\right)_{\prod_{I=1}^{5}v_{I}=1},
\end{equation}
where $\mathcal{H}(\mathscr{X})$ is the Hilbert space of $\mathbf{SQM}\left(\mathscr{X}\right)$
on $\mathbb{S}_{t}^{1}$, $F$ is the fermion number operator, and
$\beta$ is the circumference of $\mathbb{S}_{t}^{1}$. The constraint
on the fugacities $v_{I}$ decouples a $\mathrm{U}(1)\subset\mathrm{U}(1)^{5}$
symmetry that is not $\mathscr{Q}$-closed. Equivalently, $\mathscr{Z}_{\mathscr{X}}$
can be viewed as the partition function of M-theory on a fiber bundle
over $\mathbb{S}_{t}^{1}$ with fiber $\mathscr{X}$.

Based on the standard lore of Witten index, $\mathscr{Z}_{\mathscr{X}}$
only receives contributions from the ground states of $\mathbf{SQM}\left(\mathscr{X}\right)$,
which are constant modes along $\mathbb{S}_{t}^{1}$. Furthermore,
if $\mathscr{Z}_{\mathscr{X}}$ gets contributions only from supergravity
fields, including the eleven-dimensional graviton $g_{\mu\nu}$, a
Majorana gravitino $\Psi_{\mu}$, and a three-form potential $A_{\mu\nu\rho}$,
then a simple expression for $\mathscr{Z}_{\mathscr{X}}$ was derived
by Nekrasov and Okounkov \cite{Nekrasov:2014nea,Okounkov:2015spn},
\begin{equation}
\mathscr{Z}_{\mathscr{X}}^{\mathrm{SUGRA}}\left(v_{1},v_{2},v_{3},v_{4},v_{5}\right)=\mathrm{PE}\left[\mathscr{F}_{\mathscr{X}}\left(v_{1},v_{2},v_{3},v_{4},v_{5}\right)\right],
\end{equation}
where $\mathscr{F}_{\mathscr{X}}$ is computed by a geometric formula
\begin{equation}
\mathscr{F}_{\mathscr{X}}\left(v_{1},v_{2},v_{3},v_{4},v_{5}\right)=\int_{\mathscr{X}}\mathrm{ch}\left(T^{*}\mathscr{X}\ominus T\mathscr{X}\right)\wedge\mathrm{td}\left(T\mathscr{X}\right).\label{eq:FX}
\end{equation}
In particular, when $\mathscr{X}=\mathbb{C}^{5}$, we have 
\begin{equation}
\mathrm{ch}\left(T\mathbb{C}^{5}\right)=\sum_{I=1}^{5}v_{I}=\sum_{I=1}^{5}e^{x_{I}},\quad\mathrm{td}\left(T\mathbb{C}^{5}\right)=\prod_{I=1}^{5}\frac{x_{I}}{1-e^{-x_{I}}},
\end{equation}
and the formula (\ref{eq:FX}) reproduces the result obtained earlier
by Nekrasov in \cite{nekrasov2009instanton},
\begin{equation}
\mathscr{F}\left(v_{1},v_{2},v_{3},v_{4},v_{5}\right)=-\frac{\sum_{I=1}^{5}\left\llbracket v_{I}^{2}\right\rrbracket }{\prod_{I=1}^{5}\left\llbracket v_{I}\right\rrbracket },\label{eq:FC5}
\end{equation}
where we drop the subscript of $\mathscr{F}_{\mathscr{X}}$ when $\mathscr{X}=\mathbb{C}^{5}$
for simplicity. The same result was also derived in the framework
of twisted eleven-dimensional supergravity in \cite{Raghavendran:2021qbh}.
It is evident that $\mathscr{F}$ contains five singularities, $v_{I}\to1,I=1,\cdots,5$,
associated with five complex flat directions of $\mathbb{C}^{5}$.

When $\mathscr{X}$ is a product manifold of the form $\mathscr{Y}\times\mathscr{W}$,
where $\mathscr{Y}$ and $\mathscr{W}$ are respectively a Calabi-Yau
threefold and a Calabi-Yau twofolds, the holonomy group $\mathrm{SU}(5)$
is further reduced to $\mathrm{SU}(3)\times\mathrm{SU}(2)$, and we
get two extra preserved supercharges associated with the representation
$\left(\mathbf{1},\mathbf{1}\right)$ in the decomposition
\begin{equation}
\mathbf{10}_{\mathrm{SU}(5)}\to\left[\left(\mathbf{1},\mathbf{1}\right)+\left(\bar{\mathbf{3}},\mathbf{1}\right)+\left(\mathbf{3},\mathbf{2}\right)\right]_{\mathrm{SU}(3)\times\mathrm{SU}(2)},
\end{equation}
and its conjugation. In total, the number of preserved supercharges
is four, which matches that in type IIA superstring theory with all
D$6$-branes are parallel. Nevertheless, these two supercharges are
unnecessary in the definition and the computation of the index of
M-theory.

It is worthy to note that for certain types of $\mathscr{X}$, especially
when $\mathscr{X}=\mathscr{Y}\times\mathscr{W}$, the contributions
from other ingredients of M-theory cannot be neglected in order to
get a sensible index of M-theory \cite{Zotto:2021xah,Nekrasov:2021ked}.
It is beyond the scope of this paper to derive a complete formula
of $\mathscr{Z}_{\mathscr{X}}$ for general $\mathscr{X}$.

\section{Relating two counting problems \label{sec:Relation}}

So far we have discussed two counting problems, one in type IIA superstring
theory and the other in M-theory. In this section, we explore the
relation between them.

\subsection{Parallel D$6$-branes}

Let us start with the old examples in which all D$6$-branes are parallel. 

According to the duality between type IIA superstring theory and M-theory,
the D$0$-branes are lifted to the KK modes of the graviton carrying
momentum along the eleventh direction $\mathbb{S}_{R}^{1}$ of M-theory,
where $R$ is the radius of $\mathbb{S}_{R}^{1}$. Meanwhile, the
parallel D$6$-branes become KK monopoles. The geometry in the transverse
direction is an $n$-centered multi-Taub-NUT space $\mathbb{TN}_{n}$,
which approaches $\mathbb{R}^{3}\times\mathbb{S}_{R}^{1}$ at infinity,
\begin{equation}
\begin{array}{cccccc}
\mathrm{M-theory} & \mathbb{S}_{t}^{1} & \times & \mathbb{C}^{3} & \times & \mathbb{TN}_{n}\\
\uparrow & \parallel &  & \parallel\\
n\ \mathrm{D}6 & \mathbb{S}_{t}^{1} & \times & \mathbb{C}^{3}\\
 & \parallel\\
\mathrm{D}0 & \mathbb{S}_{t}^{1}
\end{array}.
\end{equation}
In the limit $R\to\infty$, the metric on $\mathbb{TN}_{n}$ reduces
to that on the $A_{n-1}$-type ALE space $\widetilde{\mathbb{C}^{2}/\mathbb{Z}_{n}}$,
which is the resolution of the orbifold $\mathbb{C}^{2}/\mathbb{Z}_{n}$.
The KK modes of the graviton and the KK monopoles carry electric and
magnetic charges under $A_{\mu}=G_{\mu,10}$, respectively. 

It is natural to expect that the instanton partition function multiplied
by a perturbative contribution should coincide with the index of M-theory
on $\mathbb{C}^{3}\times\mathbb{TN}_{n}$,
\begin{equation}
\mathcal{Z}_{n}^{\mathrm{pert}}\left(q_{1},q_{2},q_{3},q_{4}\right)\mathcal{Z}_{n}\left(q_{1},q_{2},q_{3},q_{4},p\right)=\mathscr{Z}_{\mathbb{C}^{3}\times\mathbb{TN}_{n}}\left(v_{1},v_{2},v_{3},v_{4},v_{5}\right).
\end{equation}
In the limit of infinite Taub-NUT radius $R\to\infty$, $\mathbb{TN}_{n}$
becomes $\widetilde{\mathbb{C}^{2}/\mathbb{Z}_{n}}$, and correspondingly
the $\mathrm{U}(1)$ factor in the gauge group $\mathrm{U}(n)$ decouples.
Assuming that $\mathscr{Z}_{\mathbb{C}^{3}\times\widetilde{\mathbb{C}^{2}/\mathbb{Z}_{n}}}$
can be computed exactly in the supergravity limit, one has
\begin{equation}
\mathcal{F}_{n}^{\mathrm{pert}}\left(q_{1},q_{2},q_{3},q_{4}\right)+\mathcal{F}_{n}\left(q_{1},q_{2},q_{3},q_{4},p\right)=\mathscr{F}_{\mathbb{C}^{3}\times\widetilde{\mathbb{C}^{2}/\mathbb{Z}_{n}}}\left(v_{1},v_{2},v_{3},v_{4},v_{5}\right),\label{eq:FIIAM}
\end{equation}
provided that the parameters on both sides are properly identified
\cite{nekrasov2009instanton,Nekrasov:2014nea}.

Notice that there are four independent parameters on both sides of
(\ref{eq:FIIAM}). On the type IIA superstring theory side, three
parameters encode $\mathrm{U}\left(1\right)^{3}\subset\mathrm{U}\left(1\right)^{4}$
isometries of $\mathbb{C}^{4}$ and one instanton counting parameter
labels the number of D$0$-branes. On the M-theory side, all four
parameters are fugacities associated with the $\mathrm{U}(1)^{4}$
isometries of $\mathbb{C}^{3}\times\widetilde{\mathbb{C}^{2}/\mathbb{Z}_{n}}$.
Due to the non-perturbative characteristic of the duality, one should
not expect the dictionary of the parameters to be linear.

The simplest example is when there is only one D$6$-brane, corresponding
to $\mathbb{C}^{3}\times\mathbb{C}^{2}\cong\mathbb{C}^{5}$ on the
M-theory side \cite{nekrasov2009instanton}. By comparing the formulas
(\ref{eq:F1}) and (\ref{eq:FC5}), one can easily find the identification
of the parameters,
\begin{equation}
v_{I}=\begin{cases}
q_{I}, & I=1,2,3\\
q_{4}^{\frac{1}{2}}p & I=4\\
q_{4}^{\frac{1}{2}}p^{-1} & I=5
\end{cases},
\end{equation}
and (\ref{eq:FIIAM}) is satisfied if
\begin{equation}
\mathcal{F}_{1}^{\mathrm{pert}}\left(q_{1},q_{2},q_{3},q_{4}\right)=\frac{\left\llbracket q_{4}\right\rrbracket }{\left\llbracket q_{1}\right\rrbracket \left\llbracket q_{2}\right\rrbracket \left\llbracket q_{3}\right\rrbracket }.
\end{equation}
Indeed, $\mathcal{F}_{1}^{\mathrm{pert}}$ is precisely the perturbative
contribution to the free energy of the $(6+1)$-dimensional theory
living on the D$6$-brane \cite{nekrasov2009instanton}.

The result can be generalized to the case with $n>1$ parallel D$6$-branes.
Applying the following identity \cite{Nekrasov:2018xsb},
\begin{equation}
\frac{\left\llbracket z_{1}^{n}z_{2}^{n}\right\rrbracket }{\left\llbracket z_{1}z_{2}\right\rrbracket \left\llbracket z_{1}^{n}\right\rrbracket \left\llbracket z_{2}^{n}\right\rrbracket }=\sum_{\alpha=1}^{n}\frac{1}{\left\llbracket z_{1}^{n+1-\alpha}z_{2}^{1-\alpha}\right\rrbracket \left\llbracket z_{1}^{\alpha-n}z_{2}^{\alpha}\right\rrbracket },
\end{equation}
we can decompose $\mathcal{F}_{n}\left(q_{1},q_{2},q_{3},q_{4},p\right)$
into a sum of $n$ terms,
\begin{align}
\mathcal{F}_{n}\left(q_{1},q_{2},q_{3},q_{4},p\right)= & \frac{\left\llbracket q_{1}q_{2}\right\rrbracket \left\llbracket q_{1}q_{3}\right\rrbracket \left\llbracket q_{2}q_{3}\right\rrbracket }{\left\llbracket q_{1}\right\rrbracket \left\llbracket q_{2}\right\rrbracket \left\llbracket q_{3}\right\rrbracket }\frac{\left\llbracket q_{4}^{n}\right\rrbracket }{\left\llbracket q_{4}^{\frac{n}{2}}p\right\rrbracket \left\llbracket q_{4}^{\frac{n}{2}}p^{-1}\right\rrbracket \left\llbracket q_{4}\right\rrbracket }\nonumber \\
= & \sum_{\alpha=1}^{n}\frac{\left\llbracket q_{1}q_{2}\right\rrbracket \left\llbracket q_{1}q_{3}\right\rrbracket \left\llbracket q_{2}q_{3}\right\rrbracket }{\left\llbracket q_{1}\right\rrbracket \left\llbracket q_{2}\right\rrbracket \left\llbracket q_{3}\right\rrbracket }\frac{1}{\left\llbracket q_{4}^{\alpha-\frac{n}{2}}p\right\rrbracket \left\llbracket q_{4}^{1-\alpha+\frac{n}{2}}p^{-1}\right\rrbracket }\nonumber \\
= & \sum_{\alpha=1}^{n}\mathcal{F}_{1}\left(q_{1},q_{2},q_{3},q_{4},q_{4}^{\alpha-\frac{n+1}{2}}p\right).\label{eq:decomposeFn}
\end{align}
Therefore, if we take
\begin{align}
\mathcal{F}_{n}^{\mathrm{pert}}\left(q_{1},q_{2},q_{3},q_{4}\right) & =n\mathcal{F}_{1}^{\mathrm{pert}}\left(q_{1},q_{2},q_{3},q_{4}\right)\nonumber \\
 & =\frac{n\left\llbracket q_{4}\right\rrbracket }{\left\llbracket q_{1}\right\rrbracket \left\llbracket q_{2}\right\rrbracket \left\llbracket q_{3}\right\rrbracket },
\end{align}
we have 
\begin{align}
 & \mathcal{F}_{n}^{\mathrm{pert}}\left(q_{1},q_{2},q_{3},q_{4}\right)+\mathcal{F}_{n}\left(q_{1},q_{2},q_{3},q_{4},p\right)\nonumber \\
= & \sum_{\alpha=1}^{n}\mathscr{F}\left(q_{1},q_{2},q_{3},q_{4}^{\alpha-\frac{n}{2}}p,q_{4}^{1-\alpha+\frac{n}{2}}p^{-1}\right),\label{eq:OldMatching}
\end{align}
which is equal to $\mathscr{F}_{\mathbb{C}^{3}\times\widetilde{\mathbb{C}^{2}/\mathbb{Z}_{n}}}$.
As shown in \cite{Nekrasov:2018xsb}, the $n$ terms in the sum correspond
to $n$ isolated fixed points of $\widetilde{\mathbb{C}^{2}/\mathbb{Z}_{n}}$,
with the local weights under the torus action determined by the equivariant
index of Dirac operators on $\mathbb{C}^{2}$ and $\widetilde{\mathbb{C}^{2}/\mathbb{Z}_{n}}$,
\begin{equation}
\chi\left[\widetilde{\mathbb{C}^{2}/\mathbb{Z}_{n}}\right]\left(z_{1},z_{2}\right)=\sum_{\alpha=1}^{n}\chi\left[\mathbb{C}^{2}\right]\left(z_{1}^{n+1-\alpha}z_{2}^{1-\alpha},z_{1}^{\alpha-n}z_{2}^{\alpha}\right).
\end{equation}

\subsection{Intersecting D$6$-branes}

Now we would like to generalize what we have reviewed to the case
involving intersecting D$6$-branes. Since an honest calculation from
the perspective of M-theory is not available, it is our aim to examine
whether a matching of two counting results is possible.

We first fix the perturbative contribution. If one assumes the equality
of all D$6$-branes, regardless of the spatial directions $A\in\underline{4}^{\vee}$
in the worldvolume, then the only reasonable choice is
\begin{equation}
\mathcal{F}_{\vec{n}}^{\mathrm{pert}}\left(q_{1},q_{2},q_{3},q_{4}\right)=\sum_{A\in\underline{4}^{\vee}}\frac{n_{A}\left\llbracket q_{\check{A}}\right\rrbracket }{\prod_{a\in A}\left\llbracket q_{a}\right\rrbracket }.
\end{equation}

Naively, the expression of $\mathcal{F}_{\vec{n}}\left(q_{1},q_{2},q_{3},q_{4},p\right)$
given in (\ref{eq:Fgeneral}) becomes singular not only when $q_{a}\to1$
but also when $p\to\varDelta_{\vec{n}}^{\pm\frac{1}{2}}$. Therefore,
it appears that we have six flat complex directions, which is too
many for an interpretation in terms of M-theory. We have observed
that the tetrahedron instanton partition function coincides with a
specialization of the partition function of the magnificent four model.
In fact, it is difficult to provide a M-theory reinterpretation of
$\mathcal{Z}^{\mathrm{MF}}$, partly because of the poor understanding
of the M-theory lift of D$8$- and anti-D$8$-branes. It was speculated
that a mysterious thirteen-dimensional theory should exist to reproduce
the result of $\mathcal{Z}^{\mathrm{MF}}$ \cite{Nekrasov:2017cih}.

Fortunately, there is a way to generalize (\ref{eq:decomposeFn})
so that $\mathcal{F}_{\vec{n}}$ can be decomposed into a sum of $\mathcal{F}_{1}$
with proper local identification of parameters. We first consider
the case with two stacks of D$6$-branes as an example. By applying
the following identity with $z_{1}=q_{4}^{\frac{n}{2}},z_{2}=q_{3}^{\frac{m}{2}}$,
\begin{equation}
\frac{\left\llbracket z_{1}^{2}z_{2}^{2}\right\rrbracket }{\left\llbracket z_{1}z_{2}p\right\rrbracket \left\llbracket z_{1}z_{2}p^{-1}\right\rrbracket }=\frac{\left\llbracket z_{1}^{2}\right\rrbracket }{\left\llbracket z_{1}z_{2}p\right\rrbracket \left\llbracket z_{1}z_{2}^{-1}p^{-1}\right\rrbracket }+\frac{\left\llbracket z_{2}^{2}\right\rrbracket }{\left\llbracket z_{1}^{-1}z_{2}p\right\rrbracket \left\llbracket z_{1}z_{2}p^{-1}\right\rrbracket },\label{eq:identity}
\end{equation}
we obtain
\begin{align}
\mathcal{F}_{(n,m,0,0)}\left(q_{1},q_{2},q_{3},q_{4},p\right)= & \mathcal{F}_{(n,0,0,0)}\left(q_{1},q_{2},q_{3},q_{4},q_{3}^{\frac{m}{2}}p\right)\nonumber \\
 & +\mathcal{F}_{(0,m,0,0)}\left(q_{1},q_{2},q_{3},q_{4},q_{4}^{-\frac{n}{2}}p\right)\nonumber \\
= & \sum_{\alpha=1}^{n}\mathcal{F}_{(1,0,0,0)}\left(q_{1},q_{2},q_{3},q_{4},q_{3}^{\frac{m}{2}}q_{4}^{\alpha-\frac{n+1}{2}}p\right)\nonumber \\
 & +\sum_{\beta=1}^{m}\mathcal{F}_{(0,1,0,0)}\left(q_{1},q_{2},q_{3},q_{3}^{\beta-\frac{m+1}{2}}q_{4}^{-\frac{n}{2}}p\right).
\end{align}
Accordingly, we can combine the perturbative and the instanton contributions
to get 
\begin{align}
 & \mathcal{F}_{(n,m,0,0)}^{\mathrm{pert}}\left(q_{1},q_{2},q_{3},q_{4}\right)+\mathcal{F}_{(n,m,0,0)}\left(q_{1},q_{2},q_{3},q_{4},p\right)\nonumber \\
= & \sum_{\alpha=1}^{n}\mathscr{F}\left(q_{1},q_{2},q_{3},q_{3}^{\frac{m}{2}}q_{4}^{\alpha-\frac{n}{2}}p,q_{3}^{-\frac{m}{2}}q_{4}^{1-\alpha+\frac{n}{2}}p^{-1}\right)\nonumber \\
 & +\sum_{\alpha=1}^{m}\mathscr{F}\left(q_{1},q_{2},q_{3}^{\alpha-\frac{m}{2}}q_{4}^{-\frac{n}{2}}p,q_{4},q_{3}^{1-\alpha-\frac{m}{2}}q_{4}^{\frac{n}{2}}p^{-1}\right).
\end{align}

For general $\vec{n}$, we can repeatedly use the decomposition (\ref{eq:decomposeFn})
and the identity (\ref{eq:identity}) to get
\begin{align}
\mathcal{F}_{\vec{n}}\left(q_{1},q_{2},q_{3},q_{4},p\right)= & \mathcal{F}_{\left(n_{123},0,0,0\right)}\left(q_{1},q_{2},q_{3},q_{4},q_{1}^{\frac{n_{234}}{2}}q_{2}^{\frac{n_{134}}{2}}q_{3}^{\frac{n_{124}}{2}}p\right)\nonumber \\
 & +\mathcal{F}_{\left(0,n_{124},0,0\right)}\left(q_{1},q_{2},q_{3},q_{4},q_{1}^{\frac{n_{234}}{2}}q_{2}^{\frac{n_{134}}{2}}q_{4}^{-\frac{n_{123}}{2}}p\right)\nonumber \\
 & +\mathcal{F}_{\left(0,0,n_{134},0\right)}\left(q_{1},q_{2},q_{3},q_{4},q_{1}^{\frac{n_{234}}{2}}q_{3}^{-\frac{n_{124}}{2}}q_{4}^{-\frac{n_{123}}{2}}p\right)\nonumber \\
 & +\mathcal{F}_{\left(0,0,0,n_{234}\right)}\left(q_{1},q_{2},q_{3},q_{4},q_{2}^{-\frac{n_{134}}{2}}q_{3}^{-\frac{n_{124}}{2}}q_{4}^{-\frac{n_{123}}{2}}p\right)\nonumber \\
= & \sum_{\alpha=1}^{n_{123}}\mathcal{F}_{(1,0,0,0)}\left(q_{1},q_{2},q_{3},q_{4},q_{1}^{\frac{n_{234}}{2}}q_{2}^{\frac{n_{134}}{2}}q_{3}^{\frac{n_{124}}{2}}q_{4}^{\alpha-\frac{n_{123}+1}{2}}p\right)\nonumber \\
 & +\sum_{\alpha=1}^{n_{124}}\mathcal{F}_{(0,1,0,0)}\left(q_{1},q_{2},q_{3},q_{4},q_{1}^{\frac{n_{234}}{2}}q_{2}^{\frac{n_{134}}{2}}q_{3}^{\alpha-\frac{n_{124}+1}{2}}q_{4}^{-\frac{n_{123}}{2}}p\right)\nonumber \\
 & +\sum_{\alpha=1}^{n_{134}}\mathcal{F}_{(0,0,1,0)}\left(q_{1},q_{2},q_{3},q_{4},q_{1}^{\frac{n_{234}}{2}}q_{2}^{\alpha-\frac{n_{134}+1}{2}}q_{3}^{-\frac{n_{124}}{2}}q_{4}^{-\frac{n_{123}}{2}}p\right)\nonumber \\
 & +\sum_{\alpha=1}^{n_{234}}\mathcal{F}_{(0,0,0,1)}\left(q_{1},q_{2},q_{3},q_{4},q_{1}^{\alpha-\frac{n_{234}+1}{2}}q_{2}^{-\frac{n_{134}}{2}}q_{3}^{-\frac{n_{124}}{2}}q_{4}^{-\frac{n_{123}}{2}}p\right).
\end{align}
It is remarkable that we are able to decompose$\mathcal{F}_{\vec{n}}$
into $\sum_{A\in\underline{4}^{\vee}}n_{A}$ terms of the form $\mathcal{F}_{1}$,
and $\mathcal{F}_{\vec{n}}$ can be naturally combined with $\mathcal{F}_{\vec{n}}^{\mathrm{pert}}$
which is also a sum of $\sum_{A\in\underline{4}^{\vee}}n_{A}$ terms
of the form $\mathcal{F}_{1}^{\mathrm{pert}}$,
\begin{align}
 & \mathcal{F}_{\vec{n}}^{\mathrm{pert}}\left(q_{1},q_{2},q_{3},q_{4}\right)+\mathcal{F}_{\vec{n}}\left(q_{1},q_{2},q_{3},q_{4},p\right)\nonumber \\
= & \sum_{\alpha=1}^{n_{123}}\mathscr{F}\left(q_{1},q_{2},q_{3},q_{1}^{\frac{n_{234}}{2}}q_{2}^{\frac{n_{134}}{2}}q_{3}^{\frac{n_{124}}{2}}q_{4}^{\alpha-\frac{n_{123}}{2}}p,q_{1}^{-\frac{n_{234}}{2}}q_{2}^{-\frac{n_{134}}{2}}q_{3}^{-\frac{n_{124}}{2}}q_{4}^{1-\alpha+\frac{n_{123}}{2}}p^{-1}\right)\nonumber \\
 & +\sum_{\alpha=1}^{n_{124}}\mathscr{F}\left(q_{1},q_{2},q_{1}^{\frac{n_{234}}{2}}q_{2}^{\frac{n_{134}}{2}}q_{3}^{\alpha-\frac{n_{124}}{2}}q_{4}^{-\frac{n_{123}}{2}}p,q_{4},q_{1}^{-\frac{n_{234}}{2}}q_{2}^{-\frac{n_{134}}{2}}q_{3}^{1-\alpha+\frac{n_{124}}{2}}q_{4}^{\frac{n_{123}}{2}}p^{-1}\right)\nonumber \\
 & +\sum_{\alpha=1}^{n_{134}}\mathscr{F}\left(q_{1},q_{1}^{\frac{n_{234}}{2}}q_{2}^{\alpha-\frac{n_{134}}{2}}q_{3}^{-\frac{n_{124}}{2}}q_{4}^{-\frac{n_{123}}{2}}p,q_{3},q_{4},q_{1}^{-\frac{n_{234}}{2}}q_{2}^{1-\alpha+\frac{n_{134}}{2}}q_{3}^{\frac{n_{124}}{2}}q_{4}^{\frac{n_{123}}{2}}p^{-1}\right)\nonumber \\
 & +\sum_{\alpha=1}^{n_{234}}\mathscr{F}\left(q_{1}^{\alpha-\frac{n_{234}}{2}}q_{2}^{-\frac{n_{134}}{2}}q_{3}^{-\frac{n_{124}}{2}}q_{4}^{-\frac{n_{123}}{2}}p,q_{2},q_{3},q_{4},q_{1}^{1-\alpha+\frac{n_{234}}{2}}q_{2}^{\frac{n_{134}}{2}}q_{3}^{\frac{n_{124}}{2}}q_{4}^{\frac{n_{123}}{2}}p\right).\label{eq:decomposeFgeneral}
\end{align}
Notice that we should not simplify $\mathcal{F}_{\vec{n}}$ using
$\prod_{a\in\underline{4}}q_{a}=1$ before performing the decomposition.
Remarkably, each term in (\ref{eq:decomposeFgeneral}) has only five
flat complex directions.

The existence of the decomposition (\ref{eq:decomposeFgeneral}) is
one of the most important observations made in this paper. We conjecture
that this expression gives the supergravity contribution to the index
of M-theory on a noncompact Calabi-Yau fivefold $\mathscr{X}_{\vec{n}}$,
where $\mathscr{X}_{\vec{n}}$ is obtained from the superposition
of KK monopoles. More precisely, when we lift a stack of D$6_{A}$-branes
to M-theory, the worldvolume directions $\mathbb{S}_{t}^{1}\times\mathbb{C}_{A}^{3}$
are trivially lifted, while the transverse directions $\mathbb{C}_{\check{A}}\times\mathbb{R}$
combine with the eleventh direction $\mathbb{S}_{R}^{1}$ of M-theory
into a Taub-NUT space. When there are two stacks of D$6$-branes,
the common worldvolume directions $\mathbb{S}_{t}^{1}\times\mathbb{C}^{2}$
are again trivially lifted, but we need to merge two six-dimensional
spaces $\mathbb{C}\times\mathbb{TN}$ into a single three-dimensional
space. The following is an example,
\begin{equation}
\resizebox{0.9\textwidth}{!}{$%
\begin{array}{cccccccc}
 &  &  &  & \mathbb{S}_{t}^{1}\times\mathbb{C}_{1}\times\mathbb{C}_{2} & \times\mathbb{TN}_{n_{123},n_{124}}\left(\supset\mathbb{C}_{3}\times\mathbb{C}_{4}\times\mathbb{R}\right)\\
\mathrm{M-theory} &  &  & \nearrow &  &  & \nwarrow\\
 & \mathbb{S}_{t}^{1}\times\mathbb{C}_{123}^{3} & \times & \mathbb{TN}_{n_{123}}\left(\supset\mathbb{C}_{4}\times\mathbb{R}\right) &  &  & \mathbb{S}_{t}^{1}\times\mathbb{C}_{124}^{3} & \times\mathbb{TN}_{n_{124}}\left(\supset\mathbb{C}_{3}\times\mathbb{R}\right)\\
 & \uparrow &  &  &  &  &  & \uparrow\\
n_{123}\mathrm{\ D}6_{123} & \mathbb{S}_{t}^{1}\times\mathbb{C}_{123}^{3} &  &  &  &  & n_{124}\mathrm{\ D}6_{124} & \mathbb{S}_{t}^{1}\times\mathbb{C}_{124}^{3}
\end{array}.
$%
}%
\end{equation}
It is straightforward to extend this logic to the case of four stacks
of D$6$-branes, and the result is M-theory on a ten-dimensional space,
which is the coalescence of four Calabi-Yau fivefolds $\mathbb{C}^{3}\times\mathbb{TN}$.
We then take the limit $R\to\infty$ to arrive at a Calabi-Yau fivefold
$\mathscr{X}_{\vec{n}}$. Correspondingly, the common center $\mathrm{U}(1)_{c}\subset\prod_{A\in\underline{4}^{\vee}}\mathrm{U}\left(n_{A}\right)$
in the theory $\mathcal{T}_{\vec{n},k}$ decouples. It would be helpful
if one can relate $\mathscr{X}_{\vec{n}}$ to a resolution of the
orbifold $\mathbb{C}^{5}\left/\Gamma\right.$,
\begin{equation}
\mathscr{X}_{\vec{n}}\xrightarrow{?}\widetilde{\mathbb{C}^{5}\left/\Gamma\right.},\quad\Gamma=\prod_{A\in\underline{4}^{\vee}}\mathbb{Z}_{n_{A}}.
\end{equation}

Given the relation (\ref{eq:ZMFZn}), our result also sheds light
on the non-perturbative interpretation of the magnificent four model
for special values of the Coulomb branch parameters. However, no similar
result exists for generic values of the Coulomb branch parameters.

\section{Relation with Gromov-Witten invariants \label{sec:TS}}

A fascinating correspondence between Donaldson-Thomas invariants and
Gromov-Witten invariants was discovered in \cite{maulik2006gromov1,maulik2006gromov2}.
According to this correspondence, the Donaldson-Thomas invariants
of $\mathbb{C}^{3}$, which is computed by the partition function
of D-instantons probing parallel D$5$-branes along $\mathbb{C}^{3}$
in type IIB superstring theory, is equivalent to the A-model closed
topological string theory on $\mathbb{C}^{3}$. To explore possible
generalizations of this correspondence, we consider the dimensional
reduction of our system. More precisely, we perform a T-duality of
the brane configuration shown in Table \ref{D0-D6} in type IIA superstring
theory along $\mathbb{S}_{t}^{1}$. 

The cohomological tetrahedron instanton partition function can be
deduced from its K-theoretic counterpart by introducing $q_{a}=e^{\beta\varepsilon_{a}}$
and taking the limit $\beta\to0$ with $\varepsilon_{a}$ and $p$
fixed, 
\begin{equation}
\mathcal{Z}_{\vec{n}}^{\mathrm{coh}}\left(\varepsilon_{1},\varepsilon_{2},\varepsilon_{3},\varepsilon_{4},p\right)=\lim_{\beta\to0}\mathcal{Z}_{\vec{n}}\left(e^{\beta\varepsilon_{1}},e^{\beta\varepsilon_{2}},e^{\beta\varepsilon_{3}},e^{\beta\varepsilon_{4}},p\right).
\end{equation}
The closed-form expression (\ref{eq:ZnIntro}) of $\mathcal{Z}_{\vec{n}}$
then leads to the following closed-form expression for $\mathcal{Z}_{\vec{n}}^{\mathrm{coh}}$,
\begin{equation}
\mathcal{Z}_{\vec{n}}^{\mathrm{coh}}\left(\varepsilon_{1},\varepsilon_{2},\varepsilon_{3},\varepsilon_{4},p\right)=\mathtt{PE}\left[\frac{p}{(1-p)^{2}}\sum_{A\in\underline{4}^{\vee}}n_{A}\Theta_{A}\right],
\end{equation}
where 
\begin{equation}
\Theta_{A}=-\frac{\prod_{a<b\in A}\left(\varepsilon_{a}+\varepsilon_{b}\right)}{\prod_{a\in A}\varepsilon_{a}}.
\end{equation}
In terms of the MacMahon function $\mathcal{M}_{3}\left(p\right)$
\cite{MacMohan1916},
\begin{equation}
\mathcal{M}_{3}\left(p\right)=\prod_{\ell=1}^{\infty}\left(1-p^{\ell}\right)^{-\ell}=\mathtt{PE}\left[\frac{p}{(1-p)^{2}}\right],
\end{equation}
we have 
\begin{equation}
\mathcal{Z}_{\vec{n}}^{\mathrm{coh}}\left(\varepsilon_{1},\varepsilon_{2},\varepsilon_{3},\varepsilon_{4},p\right)=\prod_{A\in\underline{4}^{\vee}}\mathcal{M}_{3}\left(p\right)^{n_{A}\Theta_{A}}.\label{eq:ZcohM3}
\end{equation}
The MacMahon function appears in the all-genus A-model topological
string partition function $Z_{\mathbb{C}^{3}}^{\mathrm{top}}$ of
$\mathbb{C}^{3}$ \cite{Bershadsky:1993cx,Gopakumar:1998ii,Okounkov:2003sp}.
In the equivariant form, 
\begin{equation}
Z_{\mathbb{C}_{123}^{3}}^{\mathrm{top}}\left(\varepsilon_{1},\varepsilon_{2},\varepsilon_{3},g_{s}\right)=\exp\left(\sum_{g=0}^{\infty}g_{s}^{2g-2}F_{g}\right)=\mathcal{M}_{3}\left(p=e^{\mathrm{i}g_{s}}\right)^{\Theta_{123}}
\end{equation}
where $g_{s}$ is the string coupling constant and $F_{g}$ is the
free energy of genus $g$. Therefore, we can express (\ref{eq:ZcohM3})
as a product of $Z_{\mathbb{C}^{3}}^{\mathrm{top}}$,
\begin{equation}
\mathcal{Z}_{\vec{n}}^{\mathrm{coh}}=\prod_{A\in\underline{4}^{\vee}}\left(Z_{\mathbb{C}_{A}^{3}}^{\mathrm{top}}\right)^{n_{A}}.\label{eq:ZcohTop}
\end{equation}
Recall that $\mathcal{Z}_{\vec{n}}^{\mathrm{coh}}$ computes the equivalent
volume of the moduli space $\mathfrak{M}_{\vec{n},k}$, which was
interpreted geometrically as a Quot scheme in \cite{Pomoni:2021hkn}.
Therefore, the identity (\ref{eq:ZcohM3}) gives a correspondence
between invariants that can be extracted from $\mathcal{Z}_{\vec{n}}^{\mathrm{coh}}$
and Gromov-Witten invariants on $\mathbb{C}_{A}^{3}$. 

In the special case $\vec{n}=\left(1,0,0,0\right)$, we can impose
the Calabi-Yau constraint $\varepsilon_{1}+\varepsilon_{2}+\varepsilon_{3}=0$
on $\mathbb{C}_{123}^{3}$, and (\ref{eq:ZcohTop}) becomes 
\begin{equation}
\mathcal{Z}_{\left(1,0,0,0\right)}^{\mathrm{coh}}=\left(\mathcal{M}_{3}\left(p\right)\right)^{\Theta_{123}}\xrightarrow{\varepsilon_{1}+\varepsilon_{2}+\varepsilon_{3}=0}\mathcal{M}_{3}\left(p\right)=Z_{\mathbb{C}_{123}^{3}}^{\mathrm{top}}.\label{eq:Zn0}
\end{equation}
Here $\mathcal{Z}_{\left(1,0,0,0\right)}^{\mathrm{coh}}$ with $\varepsilon_{1}+\varepsilon_{2}+\varepsilon_{3}=0$
computes the Donaldson-Thomas invariants for $\mathbb{C}^{3}$. Thus,
this formula leads to the usual correspondence between Donaldson-Thomas
invariants and Gromov-Witten invariants for $\mathbb{C}^{3}$. 

\section{Conclusions and open questions \label{sec:OpenQuestions}}

In this paper, we study bound states of D$0$-branes and intersecting
D$6$-branes in type IIA superstring theory. We compute the K-theoretic
tetrahedron instanton partition function and conjecture a closed-form
expression. It serves as an interpolation of the K-theoretic Donaldson-Thomas
invariants of $\mathbb{C}^{3}$ and the partition function of the
magnificent four model. What is more, we find a decomposition of the
partition function, making it possible to reinterpret the result non-perturbatively
through the index of M-theory on a noncompact Calabi-Yau fivefold.
After dimensional reducing the system, we discuss the connection between
the cohomological tetrahedron instanton partition function and Gromov-Witten
invariants.

There are several interesting open questions to be answered. It would
be desirable to find the solution in the eleven-dimensional supergravity
that describes the superposition of KK monopoles. In the limit of
large radius of M-theory circle, the solution should lead to the Calabi-Yau
fivefold $\mathscr{X}_{\vec{n}}$ that we are interested in. Then
one may try to evaluate the index of M-theory on $\mathscr{X}_{\vec{n}}$
and test our conjectured expression (\ref{eq:decomposeFgeneral}).
One may start with the simplest nontrivial situation in which there
are two stacks of intersecting D$6$-branes in type IIA superstring
theory. The corresponding M-theory is in $\mathbb{S}_{t}^{1}\times\mathbb{C}^{2}\times\mathscr{Y}$,
where $\mathscr{Y}$ is a Calabi-Yau threefold obtained from merging
two Taub-NUT spaces. The papers \cite{Xie:2017pfl,Jefferson:2018irk,Tian:2021cif,Nekrasov:2021ked}
may be relevant for this direction. 

In this work, we exclusively dealt with the case involving D$0$-branes
and D$6$-branes in the flat spacetime. It is natural to ask what
happens if the spacetime background is more general \cite{Szabo:2023ixw}
or if there are D$2$- and D$4$-branes in type IIA superstring theory.
In M-theory, these extra branes correspond to M$2$-branes and M$5$-branes,
and the computation of the index of M-theory becomes more complicated
\cite{Nekrasov:2014nea,Zotto:2021xah,Cirafici:2021yda}. 

From the mathematical viewpoint, an exciting problem is to provide
a rigorous formulation of the connection between the invariants computed
by the cohomological tetrahedron instanton partition function and
Gromov-Witten invariants. This may also turn out to be useful for
the study of Donaldson-Thomas/Pandharipande-Thomas correspondence
for Calabi-Yau fourfolds \cite{Cao:2019tnw}.

In \cite{Kontsevich:2008fj,Kontsevich:2010px}, the cohomological
Hall algebra (COHA) of Calabi-Yau threefolds was constructed. Recently
it was found that the moduli spaces of spiked instantons \cite{Nekrasov:2015wsu,Nekrasov:2016gud}
also admits an action of COHA \cite{Rapcak:2018nsl}. As there are
several resemblances between tetrahedron instantons and spiked instantons,
it would be also interesting to investigate whether there is an action
of certain generalization of COHA on moduli spaces of tetrahedron
instantons.


\acknowledgments
We are grateful to N. Nekrasov, N. Piazzalunga and M. Zabzine for discussions. 
EP and XZ are partially supported by the GIF Research Grant I-1515-303./2019. 
WY is supported by National Key R\&D Program of China with Grant NO: 2022ZD0117000 and NSFC with Grant NO: 12247103.
XZ is also funded by the Deutsche Forschungsgemeinschaft (DFG, German Research Foundation) under Germany’s Excellence Strategy - EXC 2121 "Quantum Universe" - 390833306, and  National Science Foundation of China under Grant No. 12147103.


\providecommand{\href}[2]{#2}\begingroup\raggedright\endgroup

\end{document}